\documentclass[linenumbers]{spie}  
\usepackage{lineno}

\usepackage{amsmath,amsfonts,amssymb}
\usepackage{graphicx,comment}
\usepackage[colorlinks=true, allcolors=blue]{hyperref}
\usepackage{gensymb}

\usepackage{booktabs}
\usepackage{float}
\usepackage{tabularx}  
\usepackage{subcaption}
\usepackage{caption}
\usepackage{siunitx}
\usepackage{array}
\usepackage[utf8]{inputenc}

\newcommand{\kms}{\mbox{km~s$^{-1}$}}

\title{The Ghana Radio Astronomy Observatory}

\author[a]{Emmanuel Proven-Adzri}

\author[b]{Nia Imara} 

\author[a]{Theophilus  Ansah-Narh} 

\author[c]{Wonder Sewavi}

\author[a]{Diana Klutse}

\author[a]{Evaristus Iyida}

\author[a]{Joseph Bremang Tandoh}

\author[a]{Naomi Asabre Frimpong}

\author[a]{Benedicta Woode}

\author[d]{Pieter Pretorius}

\affil[a]{Ghana Radio Astronomy Observatory, Ghana Space Science and Technology Institute, P.O. Box LG 80, Legon, Accra, Ghana}
\affil[b]{University of California, Santa Cruz, 1156 High Street, Santa Cruz, CA, USA}
\affil[c]{Department of Physics, Kwame Nkrumah University of Science and Technology, Kumasi, Ghana}
\affil[d]{South African Radio Astronomy Observatory, Liesbeek House, River Park, Gloucester Road, Mowbray, Cape Town, 7700, South Africa}

\begin{document} 
\maketitle

\begin{abstract}
The Ghana Radio Astronomy Observatory (GRAO) marks a pivotal advance in African radio astronomy through the successful transformation of a decommissioned 32\,m satellite communication antenna into a scientifically capable, VLBI-ready radio telescope. Strategically located near the equator at Kutunse, Ghana, the telescope offers nearly full-sky coverage ($-77^\circ$ to $+88^\circ$ declination), making it a valuable asset for time-domain astronomy, transient surveys, and global VLBI networks.
This work documents the technical evolution of the facility, including beam-waveguide optics, dual-polarization C-band receivers ($5$ and $6.7$ GHz), and recent backend upgrades culminating in the integration of a hydrogen maser, wideband ROACH2 system, and enhanced control and pointing infrastructure. We report early science results from high-resolution spectral-line observations of $6.7$ GHz Class II methanol masers, pulse timing of PSR J0835$-$4510 (Vela), and successful VLBI fringe detections on intercontinental baselines. Simulations and commissioning tests confirm high aperture efficiency ($\geq77\%$), low sidelobe levels, and robust time stability across the signal chain. These outcomes validate the GRAO's readiness for both standalone and networked operations.
As the first operational node in West Africa contributing to the African VLBI Network, GRAO plays a critical role in advancing the continent's participation in global radio astronomy, capacity building, and the preparatory phase of the Square Kilometre Array.

\end{abstract}

\keywords{GRAO, 32\,m radio telescope, early science, methanol masers, VLBI}

\section{INTRODUCTION}\label{sec:intro}

The Ghana Radio Astronomy Observatory (GRAO), featuring a 32\,m radio telescope, 
is Ghana's leading astronomy facility and a symbol of the country's emergence in space science and technology. A department of Ghana Space Science and Technology Institute \href{https://gssti.org}{(GSSTI)}, the GRAO is now in routine operation in Kutunse, located about 25 km north-west of Accra. The 32\,m telescope was originally constructed in 1981 as  a telecommunications dish, serving as Ghana's main node for international satellite communications. With the advent of fiber-optic technology, it became obsolete in the early $2000$s. 

Across the African continent, radio astronomy infrastructure has historically been limited, resulting in sparse geographic coverage for high-resolution interferometric science. GRAO's development forms part of a broader effort to close this gap and enable regional participation in global networks such as the African VLBI Network (AVN) and the Square Kilometre Array (SKA).

In 2011, the GSSTI--under the Ghana Atomic Energy Commission (GAEC)--along with the Square Kilometre Array, South Africa (SKA-SA), and the Hartebeesthoek Radio Astronomy Observatory (HartRAO), identified the redundant dish as a potential asset for radio astronomy and a precursor for astronomy development in Ghana.
This initiative was envisioned as a strategic effort to build national capacity and technical expertise in anticipation of the broader SKA rollout across the African continent.
The SKA South Africa, which later became the South African Radio Astronomy Observatory
\href{https://www.sarao.ac.za}{(SARAO)}, partnered with Ghana's GSSTI to initiate the necessary engineering transformation. This collaborative vision aligned closely with Ghana's long-term goals for advancing its space science capabilities. As a result, a comprehensive refurbishment programme was launched to repurpose the defunct telecommunications antenna into a scientifically capable radio telescope.

Located near the equator at 5.75$^\circ$N, the GRAO provides near-complete access to the celestial sphere 
(declinations $-77\degree$ to $+88\degree$),
offering a unique advantage for time-domain surveys, long-baseline interferometry, and the continuous monitoring of transient events. Since its inauguration in 2017, the 32\,m telescope has been operating in single-dish mode, with planned participation in Very Long Baseline Interferometry (VLBI) through the European VLBI Network (EVN) and future integration into the African VLBI Network (AVN). 

\begin{figure}
\begin{center}
    \includegraphics[width=\textwidth]{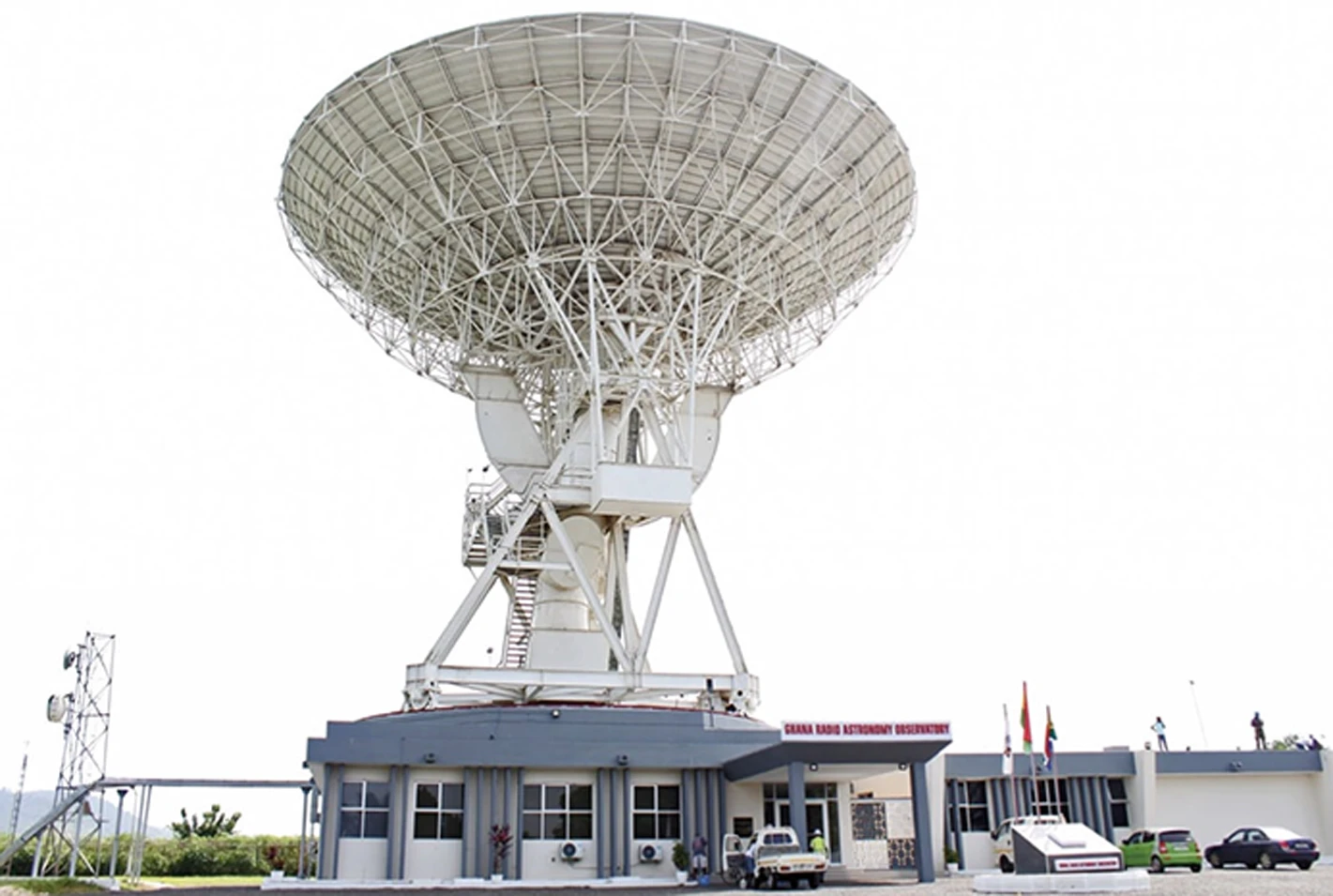}
    \caption{The Ghana Radio Astronomy Observatory}
    \label{fig:grao}
\end{center}
\end{figure}

Currently, GRAO's 32\,m dish (Fig.~\ref{fig:grao}) observes in the C-band at central frequencies of $5$\,GHz and $6.7$\,GHz, with respective bandwidths of $128$\,MHz and $400$\,MHz.

This paper provides a detailed account of the GRAO’s technical conversion from a legacy satellite communications antenna to a modern, VLBI-capable radio telescope. It outlines major instrumentation upgrades—including front-end receiver systems, timing infrastructure, and digital backends—and presents early science results that validate the facility’s operational and scientific readiness.

Key scientific targets for the GRAO include:

\begin{itemize}    \item Spectroscopy of methanol (CH$_3$OH) masers associated with high-mass star-forming regions.
    
    \item Pulsar timing to detect sudden spin-up events called glitches.
    \item Follow-up of rotating radio transits, fast radio bursts, and other transient phenomena.    \item Radio continuum flux monitoring of gamma-ray flare sources.
  
    \item In its VLBI role, the GRAO will contribute to high-resolution astrometric studies, including studying the environment of methanol masers, measuring pulsar parallaxes and proper motions, imaging active galactic nuclei (AGN) jets, observing/imaging quasars, and responding rapidly to gamma-ray burst events.
\end{itemize}

In addition, the GRAO has a comprehensive program involving public engagement and outreach
\cite{Aworka2021}. Schools regularly visit the observatory to learn about the engineering and the science research taking place at the GRAO, while the general public visits during astronomy open days. For the past 8 years, we have been training young astronomers in telescope operation and educating them in astrophysics, big-data analytics, and machine learning, turning the GRAO into a new hub for radio astronomy and the education of the next generation of astrophysicists on the African continent.

\section{Design and Observational Features of the GRAO}\label{sec:design}

The key properties and technical features of the GRAO are summarized in Table~\ref{tab:grao_spec}.
The system temperature of the telescope can be determined using a calibrated noise diode as a reference signal. The relationship is expressed as:

\begin{equation}
T_{\mathrm{sys}} = T_{\mathrm{noise\,diode}} \times \frac{P_{\mathrm{cold\,sky}}}{P_{\mathrm{noise\,diode}} - P_{\mathrm{cold\,sky}}}
\end{equation}

$T_{\mathrm{noise\,diode}}$ is the effective noise temperature of the injected noise diode signal, $P_{\rm cold -sky}$ is the measured receiver power when the telescope is pointed at a low-emission (cold) region of the sky. The difference $P_{noise\,diode} - P_{\rm cold-sky}$ isolates the response due solely to the noise diode.  With $T_{noise\,diode}$ known and the powers $P_{\rm cold-sky}$ and $P_{\rm noise\,   diode}$ measured, the system temperature $T_{\rm sys}$ can be determined directly and reliably from the above relation.

The observatory is located in Kutunse, near Accra, at geographic coordinates $5.75\degree$N latitude and $-0.305\degree$ W longitude. Its proximity to the equator enables observations across a wide declination range, $-77\degree$ to $+88\degree$, corresponding to nearly complete access to the celestial sphere and providing sky coverage of approximately $165\degree$ out of $180\degree$. 
This makes the GRAO telescope well-suited for wide-sky surveys and for participating in VLBI networks.

Located in the UTC+0 time zone, the GRAO is advantageously positioned for continuous, time-synchronized monitoring of transient astrophysical events, including AGN flares, pulsar glitches, and fast radio bursts. Its geographic location also strengthens the baseline configuration of global VLBI arrays, by serving as a central point between Southern African and European telescopes. This enhances $u$–$v$ coverage, thereby improving angular resolution, image fidelity, and network robustness. Detailed analyses of the GRAO’s contribution to VLBI array performance can be found in the following references~\cite{gaylard2014african,asabere2015radio,atemkeng2022radio}.

The telescope employs a 32\,m Cassegrain-type dual-reflector system implemented within a beam-waveguide (BWG) design. The BWG architecture incorporates four mirrors--two flat and two concave--that guide incoming signals from the sub-reflector to the receiver feed horn located approximately $20$ m below the primary dish’s vertex. The primary reflector has a focal length to diameter ($f/D$) ratio of 0.32, while the Cassegrain focus is positioned $0.84$ m above the reflector vertex. The sub-reflector, with a diameter of $2.90$ m, ensures efficient signal redirection through the optical chain. 

Structurally, the antenna is mounted on an alt-azimuth platform using a wheel-and-track system. This configuration offers mechanical simplicity, robustness, and precise pointing capability. The antenna supports slew rates of about $0.30\degree$/s in both azimuth and elevation, enabling responsive tracking of celestial objects. The receiver system operates within the C-band frequency range of $5-6.7$ GHz and includes uncooled dual-polarized feeds capable of detecting both left and right circular polarizations. These features collectively enable the telescope to support a wide range of scientific applications, from continuum imaging to spectral line observations.



\section{GRAO Status}\label{sec:status}



\subsection{Instrument status and engineering progress}

The current status of instrumentation at the GRAO reflects a deliberate and structured evolution from legacy hardware constraints to a modern, scalable, and scientifically capable architecture. During Phase I, the signal chain architecture was primarily inherited from the original Intermediate Circular Orbit (ICO) satellite system, which included waveguide components, low noise amplifiers (LNAs), and noise diode modules. Circularly polarized signals were converted to linear polarization via a polarizer, followed by an orthomode transducer (OMT) that separated them into left and right circular polarization (LCP and RCP). This process was applied independently to both the 5 GHz and 6.7 GHz initial frequency bands, resulting in four output signals interfacing with waveguide bandpass filters (BPFs) to reject out-of-band interference.
The receiver followed a two-stage heterodyne architecture, down-converting the radio frequency (RF) signal first to 2.9 GHz and subsequently to a final intermediate frequency (IF) of $600$ MHz. 

However, the backend system incorporated both a Digital Base Band Converter (DBBC) operating at 1024 Msps and a ROACH1 system sampling at 800 Msps. 
The disparity in sampling rates and IF frequencies necessitated a complex switching architecture in the third stage of the receiver.
Additional constraints included the retention of legacy ICO components to reduce cost and time, the presence of a legacy transmitter in the 5.0--6.7 GHz band that precluded simultaneous wideband reception, and a cumulative system noise temperature penalty of approximately 35 K due to insertion losses from the OMT and polarizer ($\sim 0.5$ dB).

In response to these limitations, Phase II of the upgrade initiative has introduced substantial improvements in both architecture and functionality. A single down-conversion receiver design has replaced the dual-stage configuration, thereby reducing architectural complexity. 
Control and monitoring functionalities have been embedded at each receiver stage to enable autonomous diagnostics and facilitate remote support. An adjustable attenuator has been incorporated into the noise diode module, enhancing calibration flexibility. The ROACH1 backend has been superseded by a wideband ROACH2 system, operating at 1024 Msps, identical to the DBBC, thereby eliminating the need for dual IF chains and enabling a common IF output. This architectural unification permits wider RF bandwidths, constrained only by the front-end ICO limitations.

Parallel to electronic improvements, structural and mechanical upgrades have also been a major focus under Phase II. The transition from a stationary pintle post to a rolling element azimuth bearing addresses previous slippage and lateral movement concerns, improving the mechanical reliability of the 32\,m antenna. Concurrently, enhancements to the control system have led to substantial gains in operational precision. Notably, the implementation of the Azimuth Angle Encoding System has yielded marked improvements in azimuthal measurement accuracy, aligning with AVN specifications.

A critical development is the design and deployment of the Drive Adapter System (DAS), engineered specifically for AVN compatibility. The DAS eliminates redundant transfer cases, reduces maintenance overhead, and improves drive system performance. Its successful integration at both the azimuth and elevation axes represents a milestone achievement, effectively aligning the GRAO mechanical interface with broader AVN infrastructure standards. 
Foundational geodetic improvements have also been achieved through the establishment of pillars for the Local Ground Control Network (LGCN) at the Kutunse site. These serve as reference points for integrating survey-grade instrumentation necessary for high-precision Antenna Reference Point (ARP) measurements. Environmental considerations were incorporated throughout construction to ensure long-term stability and accuracy.

On the software front, updates to Station Monitoring and Control (SCAM) which consists of Station Controller Software (SCS) and Infrastructure Controller Software (ICS) have been implemented to streamline operations. This includes revisions to the SCS and integration with auxiliary systems, such as the Maser Building Controller (MBC), Diesel Generator Controller (DGC), and the Environmental Monitoring System (EMS). These updates have enhanced system automation, monitoring, and robustness, key prerequisites for long-duration VLBI campaigns.

One of the final and most critical engineering undertakings involves the Timing and Frequency Reference System. This includes the successful delivery and installation of a Hydrogen Maser, whose stability is essential for both single-dish and VLBI operations. Engineering activities have focused on preparing a controlled environment at the GRAO site for housing the maser, including installation of cable trays, earthing strips, anti-static flooring, and temperature control systems. Vibration testing on the pedestal and measurement of cable path lengths were conducted to ensure signal integrity and phase stability.

At present, the upgraded receiver has been installed and is operational, albeit with minor issues that are currently under review. The Hydrogen Maser has also undergone remote test runs, demonstrating acceptable stability metrics. With approximately $98\%$ of Phase II implementation completed, the project is now in the final testing stage, which is centered on validating maser synchronization and clock stability. The forthcoming milestone will be the commencement of soft VLBI observations, marking a key transition into routine scientific operations.

\subsection{Beam pattern characterization}

The beam pattern of a radio telescope defines its directional sensitivity and is fundamental to determining angular resolution, calibration fidelity, and susceptibility to off-axis source contamination. For the GRAO 32-m dish, we simulated the beam response assuming an idealized circular aperture, uniformly illuminated, and corrected for surface roughness using Ruze’s approximation. The baseline theoretical pattern is governed by the Airy diffraction model, expressed as Eq.~\eqref{eq:bs}:

\begin{equation} \label{eq:bs}
G_{\text{ideal}}(\theta) = \left[ \frac{2J_1\left(\dfrac{\pi D \sin\theta}{\lambda}\right)}{\dfrac{\pi D \sin\theta}{\lambda}} \right]^2,
\end{equation}

\noindent where $J_1$ is the first-order Bessel function of the first kind, $D = 32$ m is the dish diameter, $\lambda$ is the observing wavelength, and $\theta$ is the angular offset from boresight in degrees. This ideal response is attenuated by the Ruze factor to account for surface irregularities, given by Eq.~\eqref{eq:rs}:

\begin{equation} \label{eq:rs}
\eta_{\text{Ruze}} = \exp\left[-\left( \frac{4\pi \sigma}{\lambda} \right)^2 \right],
\end{equation}

\noindent where $\sigma = 1.88$ mm is the RMS surface error as stated in the GRAO technical specification sheet.

The resulting beam gain patterns at $5.0$ GHz and $6.7$ GHz are shown in the left panel of Fig.2 .
The modeled profiles are normalized to $0$ dB peak gain and reveal a series of well-structured sidelobes and nulls consistent with ideal diffraction behavior.
Both bands conform to the expected HPBW: $0.11^\circ$ at $5$ GHz and $0.09^\circ$ at $6.7$ GHz. Sidelobe levels in the simulation also align with specification constraints, registering approximately $-15.21$ dB and $-15.15$ dB, respectively, which is indicative of effective edge taper and low spillover.

To validate these results, we compared our simulation with the electromagnetic and observational beam profiles reported by \cite{venter2018electromagnetic}, who utilized \texttt{GRASP} physical optics modeling as well as commissioning scans of bright radio sources, including Cygnus A and Taurus A.
In their result, the \texttt{GRASP}-derived patterns for the nominal subreflector alignment exhibit nearly identical HPBW and null spacing compared to our simulated result.
Notably, their study also investigated subreflector offsets of $\pm 2.25$ cm in the zenith direction (green and magenta curves), which introduced beam asymmetries and elevated sidelobe levels, effects absent in our current model, which assumes static mechanical alignment. This corroboration affirms that, under nominal alignment, the GRAO antenna exhibits beam characteristics consistent with both theoretical expectations and electromagnetic analysis.
 These patterns can serve as prior models in direction-dependent calibration schemes and are particularly relevant for pulsar timing, continuum imaging, and transient searches requiring high beam shape stability \cite{perley2017accurate}.

 \begin{figure}[H]
 	\centering
 	\includegraphics[width=\textwidth]{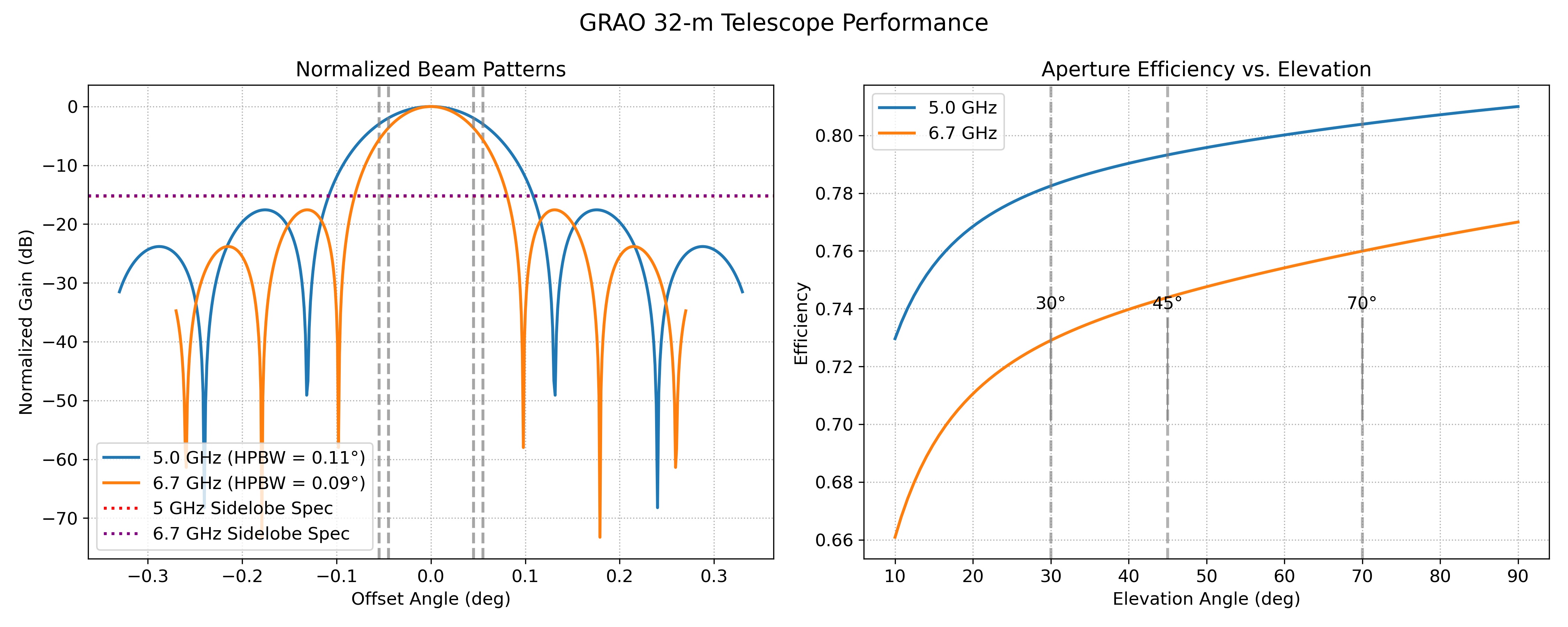}
 	\caption{Antenna performance simulation for the GRAO.
     Left panel: Normalized beam gain patterns at $5.0$ GHz and $6.7$ GHz, including sidelobe specifications (dashed lines). The simulations assume ideal optical alignment and a surface RMS error of $1.88$ mm.
     Right panel: Aperture efficiency as a function of elevation angle, incorporating surface deformation, atmospheric opacity, and pointing losses. Dashed vertical lines denote typical operational elevations ($30^\circ$, $45^\circ$, and $70^\circ$). Peak efficiencies match design specifications at zenith, while degradation at low elevation reflects expected structural and atmospheric constraints.}
     \label{fg:GRAO_efficiency_final_spec}
 \end{figure}

\subsection{Elevation dependence of aperture efficiency}

The aperture efficiency, $\eta$, quantifies the fraction of incident electromagnetic energy effectively coupled into the receiver system. While the peak efficiency at zenith ($\eta_0$) is specified as 0.81 and 0.77 for 5 GHz and 6.7 GHz respectively, this value degrades with elevation due to structural deformation, atmospheric opacity, and pointing inaccuracies. These effects are especially pertinent in tropical environments such as Ghana, where thermal expansion and humidity can further compound performance degradation.

The elevation-dependent aperture efficiency is modeled as Eq.~\eqref{eq:el1}:

\begin{equation} \label{eq:el1}
\eta(\mathrm{EL}) = \eta_0 \cdot \dfrac{ \eta_{\text{Ruze}}(\mathrm{EL}) \cdot \eta_{\text{atm}}(\mathrm{EL}) \cdot \eta_{\text{pointing}} }{ \eta_{\text{total}}(\mathrm{EL}=90^\circ) },
\end{equation}

\noindent where each term is normalized to its zenith value to enforce $\eta(90^\circ) = \eta_0$. 
Ruze efficiency incorporates elevation-dependent surface deformation via Eq.~\eqref{eq:el2}:

\begin{equation} \label{eq:el2}
\sigma_{\text{eff}}(\mathrm{EL}) = \sigma \cdot \left[ 1 + \delta \cdot \cos(\mathrm{EL}) \right],
\end{equation}

\noindent where $\delta = 0.06$ models a 6\% flexure increase at low elevation, consistent with large equatorial mounts. Atmospheric losses are captured using Eq.~\eqref{eq:el3}:

\begin{equation} \label{eq:el3}
\eta_{\text{atm}}(\mathrm{EL}) = \exp\left( -\frac{\tau}{\sin(\mathrm{EL})} \right),
\end{equation}

\noindent with zenith opacities of $\tau = 0.018$ (5 GHz) and $\tau = 0.025$ (6.7 GHz) representative of Ghana’s mid-humidity observing conditions. Pointing loss is treated using a Gaussian beam model as presented in Eq.~\eqref{eq:el4}:

\begin{equation} \label{eq:el4}
\eta_{\text{pointing}} = \exp\left[ -2.77 \left( \frac{ \theta_{\text{err}} }{ \theta_{\text{beam}} } \right)^2 \right],
\end{equation}

\noindent where $\theta_{\text{err}} = 0.008^\circ$ (5 GHz) and $0.012^\circ$ (6.7 GHz), and the beamwidth $\theta_{\text{beam}} = \dfrac{70 \lambda}{D}$ reflects the telescope’s resolving limit.

The right panel of Fig.~\ref{fg:GRAO_efficiency_final_spec} illustrates the efficiency model across $10^\circ$--$90^\circ$ elevation.
At low elevation ($\sim 10^\circ$), the model yields efficiencies of $0.73$ ($5$ GHz) and $0.66$ ($6.7$ GHz), increasing to $0.82$ and $0.75$, respectively at the zenith.
These trends are physically intuitive: tilting toward the horizon increases structural sag and atmospheric path length, both of which reduce effective gain.

\subsection{GRAO data quality and operational performance}\label{sec:performance}

The GRAO has faced several challenges during its commissioning and early science operations.
These include mechanical pointing uncertainties inherited from its legacy INTELSAT structure, elevated system temperatures ($\sim 110$ K), and delays resulting from the COVID-19 pandemic.
 The current configuration of the observatory, limited to C-band reception with a basic ROACH2-based back-end, has restricted its usable bandwidth, sensitivity, and spectral resolution. In parallel, the initial lack of mature calibration pipelines and limited local technical capacity further constrained operational throughput.

Several upgrade pathways are underway to address these issues. These include the integration of high-resolution encoders, improved tracking models, and upgraded timing infrastructure to refine pointing accuracy. Prospective hardware improvements, such as cryogenic receivers and VLBI-optimized digital back-ends, aim to reduce system temperature and increase compatibility with the broader AVN. 
The deployment of RFI monitoring instrumentation and targeted bandpass filtering has begun to mitigate persistent interference from terrestrial and satellite-based emitters operating near the telescope's observing frequencies, a particular concern in GRAO’s semi-urban setting.

On the data handling side, automated reduction workflows are being developed to support routine calibration, flagging, and imaging, with an emphasis on reproducibility and readiness for VLBI pipelines. Human capital development spearheaded by partnerships with UCSC, the \href{https://www.dara-project.org}{DARA} project, and SARAO has resulted in a growing pool of trained engineers and scientists now contributing directly to observatory operations.

The successful reproduction of the performance of the specification level at the zenith reinforces confidence in the electromagnetic design and mechanical integrity of the antenna.
However, the simulation results also reveal a measurable loss in aperture efficiency at lower elevations, highlighting the importance of elevation-aware scheduling and real-time calibration strategies. For instance, high-sensitivity observing campaigns may prioritize transits near zenith to minimize losses from gravitational flexure and atmospheric attenuation. These findings offer both operational guidance and strategic foresight.

More broadly, the modeling framework developed here provides a blueprint for evaluating similar telescopes across the AVN, particularly in tropical settings where environmental loading can degrade system performance.
The ability to quantify such degradation supports harmonized diagnostics, cross-facility benchmarking, and informed scheduling all critical for the long-term scientific return of a geographically distributed interferometric array.

\section{Science Results}\label{sec:results}

Science operations and commissioning are now occurring concurrently at GRAO. The results of astronomical sources observed intermittently between August 2023 and April 2025 are presented here. 
These sources were restricted to declinations above -53 degrees. Expected sensitivity during 30 minutes on-source integration in $0.022~\kms$
channels assuming a Tsys of 90 K reaching noise levels of 0.014 Jy per channel.

Observation modes are tuned for broadband continuum, narrow spectral lines, and pulsars.
The bandwidth for continuum observations at 5 GHz is 128 MHz and at 6.7 GHz is $400$ MHz. 
The narrow band set-up has 2 MHz bandwidth and 4096 channels across to achieve a total bandwidth per channel of 0.488 kHz.  This provides high spectral resolution to fully sample the methanol masers. This is sufficient to monitor the full extent of the different velocity
components for each methanol maser source, which is usually more than 89 \kms.
Position switching observing mode was used to account for the sky background and observed in both left and right circular
polarizations. Together with regular pointing and flux calibrations, each source requires about 30 minutes of observation time.



\subsection{Methanol maser observations}
As part of its early science and commissioning phase, the GRAO conducted spectral line observations of Class II methanol masers at 6.7 GHz using its 32\,m single-dish telescope. 

\begin{figure} 
 \begin{center}
     \includegraphics[width=\textwidth]{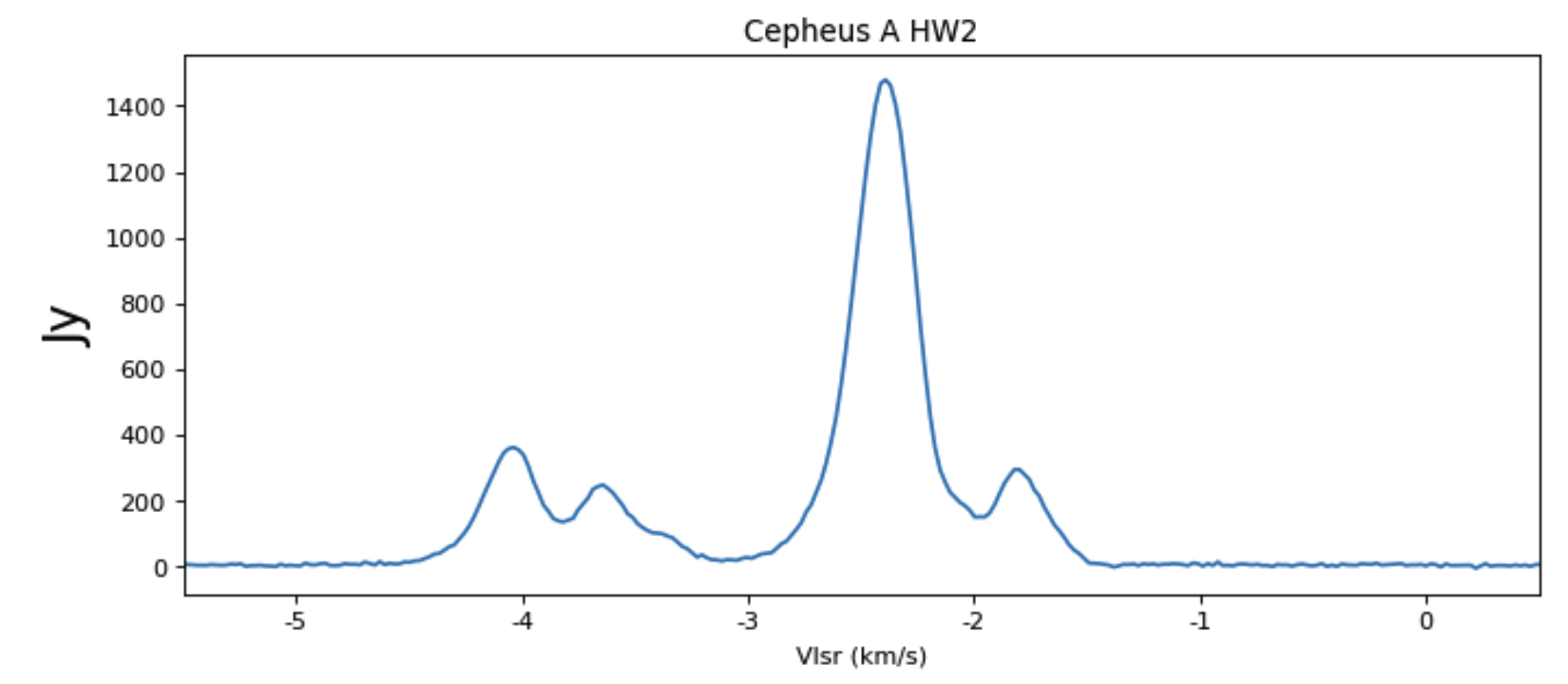}
 \caption{Methanol maser observation towards Cepheus A HW2 with the Kutunse radio telescope in Ghana on 04 March 2025. The axes are labeled for flux density (Jy) and Velocity of local standard of rest.}
  \label{fig:cepheusA}
 \end{center}
 \end{figure}

The system achieved spectral resolutions of 0.488\,kHz across a 2\,MHz bandwidth, enabling precise velocity profiling of maser features spanning over 89\,km\,s$^{-1}$.
Observations targeted key high-mass star-forming regions, including Cepheus A HW2, G009.62$+$00.20E, and G006.795$-$0.257. 
The methanol maser in Cepheus A HW2, shown in Fig.~\ref{fig:cepheusA}, exhibited an increase in flux density and the emergence of a secondary peak, hinting at dynamic outflow structures compared with previous observations~\cite{CepheusA_Durjasz_2022A&A}.
G009.62$+$00.20E, a well-studied periodic methanol maser source, was regularly used during system testing and revealed a slight decrease in flux relative to earlier data.
G006.795$-$0.257, observed near W28, a supernova remnant, exhibited a strong primary feature at 142\,Jy.
These results validate GRAO’s capabilities in high-resolution maser science and lay the groundwork for long-term monitoring of massive star-forming regions from Ghana.

\subsection{PSR J0835-4510 Vela pulsar}

\begin{figure} 
 \begin{center}
     \includegraphics[width=\textwidth]{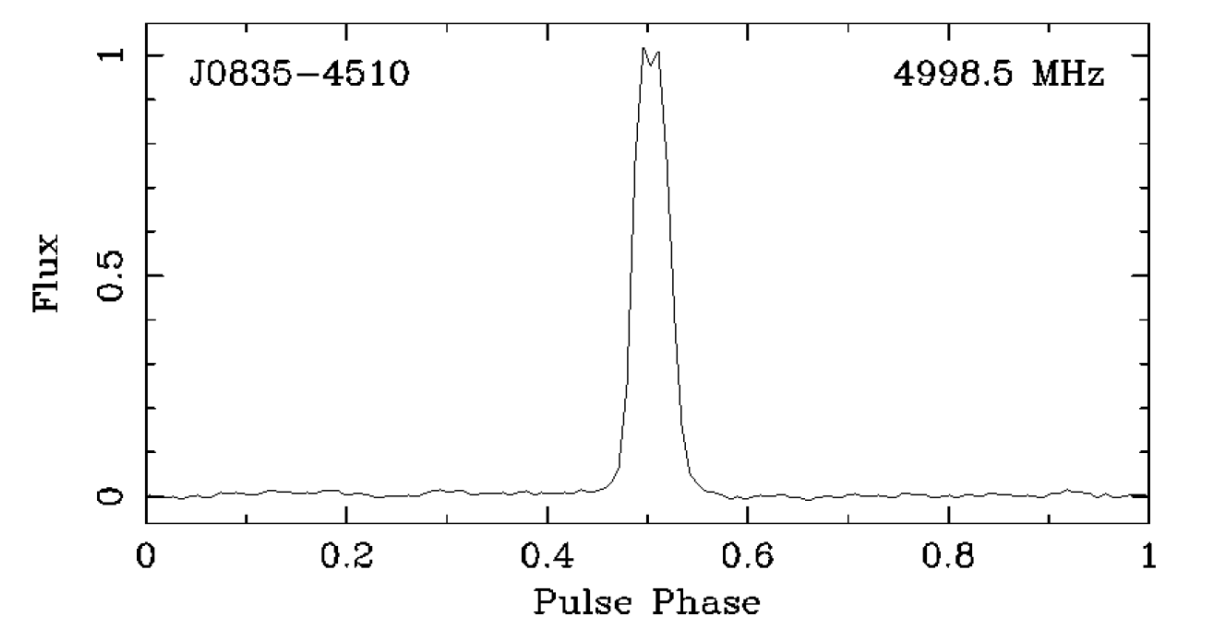}
 \caption{A 61-minute pulse profile of PSR J0835-4510 Vela pulsar taken with the GRAO telescope in August 2018.}
  \label{fig:vela_grao}
  \end{center}
 \end{figure}

An observation of Vela pulsar PSR J0835-4510, at a distance of $\sim 300$ pc, is presented in Fig.~\ref{fig:vela_grao}. The profile is that of a 61-minute observation at 5 GHz, with a bandwidth of 75 MHz observed August 2018.
Standard pulsar timing techniques to measure the times of arrival (TOAs) of individual pulses were employed. The rotational behavior of Vela was then characterized. This preliminary result of its pulse profile represents the variation in observed flux over a full rotational cycle, typically normalized from phase 0 to 1.

\subsection{Quantitative Sensitivity Check: Radiometer Equation Estimate for Vela}

To quantitatively validate the detection of the Vela pulsar, we first define the system equivalent flux density (SEFD), which characterizes the intrinsic sensitivity of the telescope:

\begin{equation}
\mathrm{SEFD} = \frac{2 k T_{\mathrm{sys}}}{A_{\mathrm{eff}}} \times 10^{26}~\mathrm{Jy},
\end{equation}

where $T_{{sys}}$ is the system temperature and $A_{\mathrm{eff}}$ is the effective collecting area of the telescope.  Using the SEFD, the expected signal-to-noise ratio (S/N) for pulsar observations is given by the radiometer equation

\begin{equation}
{S/N} = \frac{S \, \sqrt{n_p \, B \, t_{\mathrm{int}}}}{\mathrm{SEFD}} \, \sqrt{\frac{1 - d}{d}},
\end{equation}

 where \(S = 1.00~\mathrm{mJy}~(0.001~\rm Jy)\) is the peak flux density of Vela pulsar (see Fig. 4), \(n_p\) is the number of summed polarizations, \(B\) is the observing bandwidth (Hz), \(t_{\rm int}\) is the integration time (s), and \(d\) is the pulse duty cycle. The factors \(n_p\) and \(d\) account for the reduction in noise from summing polarizations (\(\sqrt{n_p}\)) and the effect of the pulsar's duty cycle (\(\sqrt{(1-d)/d}\)) on the effective signal-to-noise ratio. With the telescope parameters adopted here,  system temperature \(T_{\rm sys} = 91~\rm K\) at 5 GHz, effective area \(A_{\rm eff} = 619~\rm m^2\) — the system equivalent flux density is calculated as \(\mathrm{SEFD} \approx 406~\rm Jy\). For the pulsar tracking observation, the integration time was \(t_{\rm int} = 61~\rm minutes\). Assuming a representative duty cycle \(d = 0.05\), the estimated signal-to-noise ratio is scaled with the square root of the effective bandwidth.

Table~\ref{tab:vela_snr} summarizes the estimated S/N values for a range of
bandwidths. Even for modest bandwidths (100--200 MHz), the  S/N is comfortably above unity which is consistent with the observed detection.


The estimated signal-to-noise ratios (S/N~$\approx$~9--18 for 100--400~MHz bandwidth) confirm that the Vela detection is quantitatively consistent with the telescope’s sensitivity and the expected flux density at 5 GHz. This provides a useful validation of the receiver’s performance and confirms the telescope’s capability for high-frequency pulsar studies.  Given that no pulse residuals or multi-epoch timing analysis are yet presented,  we omit detailed timing claims at this stage. Future work will include coherent folding of multiple transits and time-of-arrival analyses to establish full timing capability.

\subsection{VLBI fringe observations}

\begin{figure} 
 \begin{center}
     \includegraphics[width=0.8\textwidth]{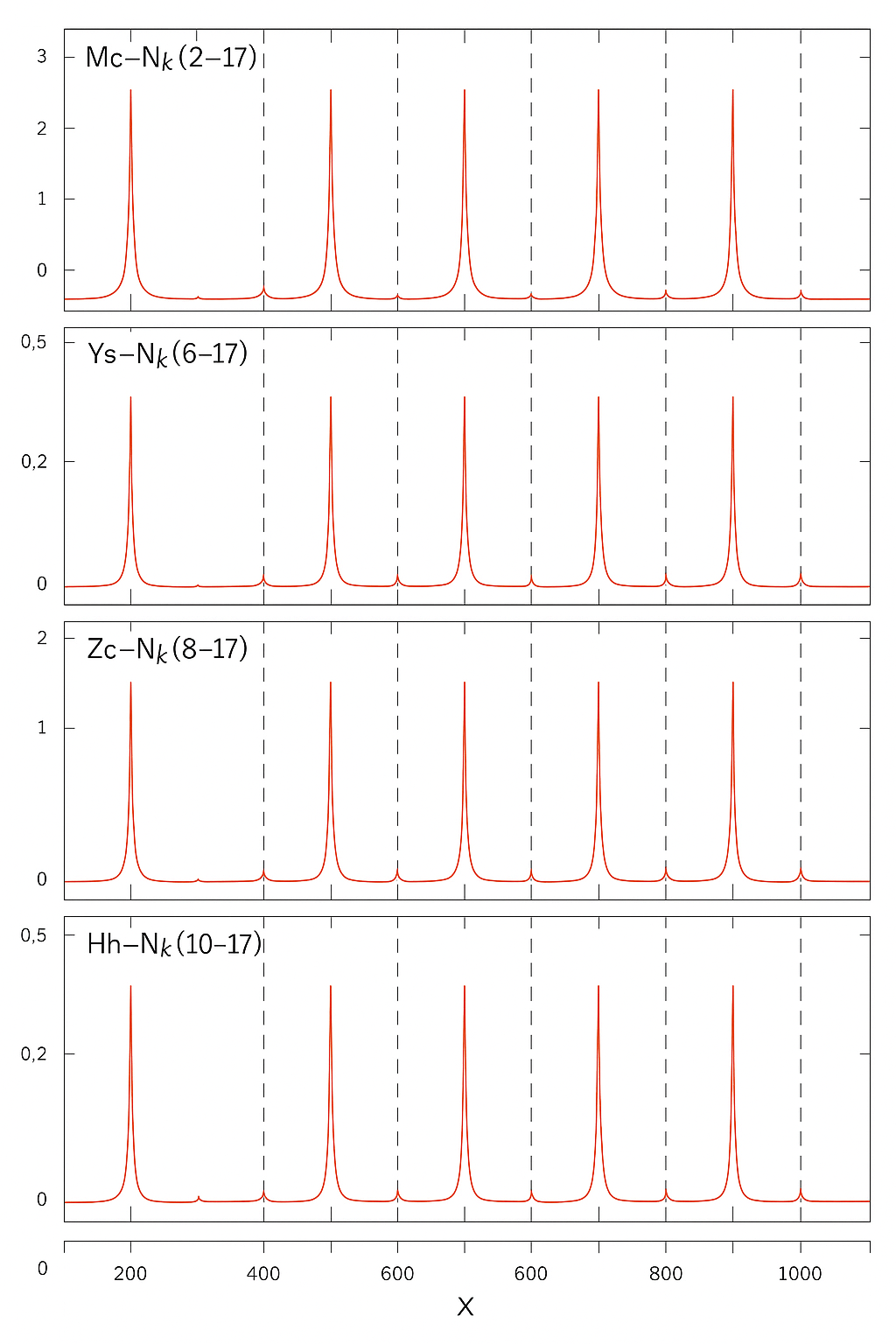}
    
 \caption{VLBI fringe tests to the GRAO telescope (Kutunse) on baselines to the telescopes in Medicina, Yebes, Zelenchukskaya and Hartebeesthoek\cite{Gurvits2021}.}
  \label{fig:vlbi_fringe}
  \end{center}
 \end{figure}

A VLBI fringe test observation was conducted in 2017 with the C-band receiver\cite{Gurvits2021}.  
Fig.~\ref{fig:vlbi_fringe} presents clear VLBI fringes obtained at JIVE (The Joint Institute for VLBI ERIC) on baselines between Kutunse and telescopes in Medicina, Yebes, Zelenchukskaya, and Hartebeesthoek located in Europe and South Africa. JIVE is a research infrastructure providing central support to the European VLBI Network (EVN). The VLBI fringe test is one of the most critical steps in verifying that the newly commissioned GRAO radio telescope can successfully participate in VLBI science observations. It confirms that the various components of the signal chain (receiver, DBBC, Mark5B recording system, clock) are in sync with the VLBI correlator at JIVE and other telescopes in the array within nanoseconds. The local oscillator and location of the telescope are stable and accurate. A new fringe test is expected by the end of 2025.

\section{Developmental Goals}

The GRAO plays a strategic role in advancing radio astronomy in West Africa and beyond. This section outlines key development goals that guide its operations, infrastructure upgrades, and science mission. These goals span from single-dish survey science to global VLBI participation, pulsar timing, and human capacity development. Ongoing and future projects are aligned to ensure GRAO’s relevance in global astronomy research and its readiness for the SKA era.

\subsection{Single-dish science}
GRAO’s 32\,m single-dish C-band telescope is developed to lead high-cadence Galactic Plane surveys, kinematic studies, and transient monitoring that exploits every advantage of a stand-alone dish: its wide beam lets it sweep large sky swaths rapidly; its full-beam sensitivity to diffuse emission will trace extended H II regions, supernova remnants shells and AGN lobes that VLBI resolves out; flexible, local scheduling enables nightly monitoring of 6.7 GHz methanol masers and blazar flares; while on-the-fly triggers capture fast radio transients within minutes. Moderate 2 MHz/400 MHz spectral modes provide simultaneous line and continuum data for variability and polarization studies, doubling science return. Single-dish science observation will supply VLBI arrays with up-to-date flux densities and preselection of flaring sources, all while serving as a cost-effective training ground where Ghanaian students learn observation planning and scheduling, survey design, RFI mitigation and Python pipelines--advancing both regional capacity and global time-domain astrophysics. Masters students supported by UCSC and DARA have begun monitoring campaigns in methanol masers and AGN studies.

\subsection{VLBI observations}
The 32\,m instrument seeks to become a regular participant in global VLBI experiments by leveraging its equatorial location to improve uv-coverage, particularly for north–south baselines within the EVN, future African VLBI Network (AVN) and other global arrays. After a successful first VLBI fringe test with the EVN in 2017, GRAO is working with JIVE to conduct a second, more advanced test by the end of 2025, while also preparing for a soft VLBI test with Hartebeesthoek. Contributing to fringe tests and calibration for global arrays, these efforts develop expertise in VLBI instrumentation, time synchronization, and correlation analysis. These will also provide hands-on training opportunities for students and engineers in Ghana, strengthen regional partnerships, and ensure that Ghana plays an active role in the global radio astronomy community as VLBI science continues to expand in both scope and impact.

\subsection{Pulsar Timing and Discovery}
GRAO is geared to study the evolution, variability of the pulse profiles and derive a spectral index for known pulsars. With repetitive scans the 32\,m radio telescope may also be used to find new intermittent pulsars missed by other search programs. The GRAO could also participate in the long-term timing of pulsars in both the C-band and (future) L-band.

\section{Summary}

The Ghana Radio Astronomy Observatory has established itself as a competitive and collaborative node in the global radio astronomy community. Its 32\,m telescope, with a shaped Cassegrain beam-waveguide design and dual-polarization C-band receivers, offers high surface accuracy (1.88 mm RMS), aperture efficiency exceeding 77 percent, and tracking performance of 0.05° RMS, consistent with specifications. Strategically located near the equator in the UTC+0 time zone, GRAO provides unmatched sky coverage and time-synchronized access to northern and southern celestial targets, a critical asset for transient monitoring and VLBI. The facility is equipped with a hydrogen maser, precision angle encoders, and antenna control drive systems, all of which ensure timing stability and mechanical accuracy essential for VLBI science. The successful observation of Cepheus A demonstrates the telescope’s readiness to participate in target of opportunity observations within its declination range.

Early observations of bright and weak methanol masers, the Vela pulsar, and successful VLBI fringe detections validate the telescope’s scientific performance. As all Phase II upgrades near completion, GRAO is technically and scientifically prepared for routine single-dish operations, further VLBI participation, and targeted science campaigns, including methanol maser monitoring, pulsar timing, and transient detection. It continues to support undergraduate and postgraduate training, regional capacity development, and Africa’s broader integration into the global SKA-VLBI ecosystem.

\section{Disclosures}
The authors declare that there are no conflicting interests in this research work. 

\section{Code and Data}
The data utilized in this study were obtained from the Ghana Radio Astronomy Observatory. Data are available from the authors upon request, and with permission from the Ghana Radio Astronomy Observatory (grao@gaec.gov.gh).

\section{Acknowledgments}
This research was fully supported under the collaborative framework established by the Memorandum of Understanding between the Ghana Atomic Energy Commission (GAEC), acting through the Ghana Space Science and Technology Institute (GSSTI), and the Regents of the University of California on behalf of its Santa Cruz Campus (UCSC). The collaboration focuses on advancing joint research and educational activities in astronomy and astrophysics, with particular emphasis on the development and scientific utilisation of the Ghana Radio Astronomy Observatory (GRAO). Funding and institutional support for this project were provided through the UCSC Santa Cruz-GRAO Astronomy Development Project. The authors gratefully acknowledge the logistical and technical assistance of the GAEC/GSSTI and the UCSC Department of Astronomy and Astrophysics.

The authors are also grateful for the  Leverhulme Royal Society UK Africa awards, DARA UK's International Science Partnership Fund via STFC, and NRF/SARAO for their contributions to astronomy research and development in Ghana.

We would also like to acknowledge and thank the referees, whose comments helped to improve this paper.

Data for this research work was obtained from the 32\,m radio telescope at the Ghana Radio
Astronomy Observatory (GRAO). The GRAO telescope is operated by the Ghana Radio Astronomy
Observatory, which is a facility of the Ghana Space Science and Technology Institute, Ghana Atomic
Energy Commission an agency of the Ministry of Environment Science Technology and Innovation. The authors are also grateful for the help of Dora Alakiya, Joel Kwakye, Henry Ayuk Agbor and Jimmy Nanewortor in making observations a success.

\bibliography{report} 

@article{perley2017accurate,
  title={An accurate flux density scale from 50 MHz to 50 GHz},
  author={Perley, Richard A and Butler, Bryan J},
  journal={The Astrophysical Journal Supplement Series},
  volume={230},
  number={1},
  pages={7},
  year={2017},
  publisher={IOP Publishing}
}

@inproceedings{venter2018electromagnetic,
  title={Electromagnetic analysis and preliminary commissioning results of the shaped dual-reflector 32-m Ghana radio telescope},
  author={Venter, M and Bolli, Pietro},
  booktitle={IOP Conference Series: Materials Science and Engineering},
  volume={321},
  number={1},
  pages={012003},
  year={2018},
  organization={IOP Publishing}
}

@INPROCEEDINGS{Gurvits2021,
       author = {{Gurvits}, Leonid I. and {Beswick}, Robert and {Hoare}, Melvin and {Njeri}, Ann and {Blanchard}, Jay and {Sharpe}, Carla and {Tiplady}, Adrian and {de Witt}, Aletha},
        title = "{High-resolution radio astronomy: An outlook for Africa}",
    booktitle = {Nuclear Activity in Galaxies Across Cosmic Time},
         year = 2021,
       editor = {{Povi{\'c}}, Mirjana and {Marziani}, Paola and {Masegosa}, Josefa and {Netzer}, Hagai and {Negu}, Seblu H. and {Tessema}, Solomon B.},
       series = {IAU Symposium},
       volume = {356},
        month = jan,
        pages = {137-142},
          doi = {10.1017/S1743921320002744},
archivePrefix = {arXiv},
       eprint = {2001.04576},
 primaryClass = {astro-ph.IM},
       adsurl = {https://ui.adsabs.harvard.edu/abs/2021IAUS..356..137G}
}

@ARTICLE{Aworka2021,
       author = {{Aworka}, R. and {Proven-Adzri}, E. and {Ansah-Narh}, T. and {Koranteng-Acquah}, J. and {Aggrey}, E.},
        title = "{Using Ghana's 32-m radio telescope to promote astronomy outreach}",
      journal = {Nature Astronomy},
         year = 2021,
        month = dec,
       volume = {5},
        pages = {1199-1202},
          doi = {10.1038/s41550-021-01555-1},
       adsurl = {https://ui.adsabs.harvard.edu/abs/2021NatAs...5.1199A}
}

@misc{gaylard2014african,
  title={An African VLBI Network of radio telescopes},
  author={Gaylard, MJ and Bietenholz, MF and Combrinck, L and Booth, RS and Buchner, SJ and Fanaroff, BL and MacLeod, GC and Nicolson, GD and Quick, JFH and Stronkhorst, P and others},
  journal={arXiv preprint arXiv:1405.7214},
  year={2014}
}

@article{atemkeng2022radio,
  title={Radio Astronomical Antennas in the Central African Region to Improve the Sampling Function of the VLBI Network in the SKA Era?},
  author={Atemkeng, Marcellin and Okouma, Patrice and Maina, Eric and Ianjamasimanana, Roger and Zambou, Serges},
  journal={Sensors},
  volume={22},
  number={21},
  pages={8466},
  year={2022},
  publisher={MDPI}
}

@misc{asabere2015radio,
  title={Radio astronomy in Africa: the case of Ghana},
  author={Asabere, Bernard Duah and Gaylard, Michael and Horellou, Cathy and Winkler, Hartmut and Jarrett, Thomas},
  journal={arXiv preprint arXiv:1503.08850},
  year={2015}
}

@ARTICLE{CepheusA_Durjasz_2022A&A,
       author = {{Durjasz}, M. and {Szymczak}, M. and {Olech}, M. and {Bartkiewicz}, A.},
        title = "{Discovery of recurrent flares of 6.7 GHz methanol maser emission in Cepheus A HW2}",
      journal = {\aap},
         year = 2022,
        month = jul,
       volume = {663},
          eid = {A123},
        pages = {A123},
          doi = {10.1051/0004-6361/202243552},
archivePrefix = {arXiv},
       eprint = {2205.08759},
 primaryClass = {astro-ph.GA},
       adsurl = {https://ui.adsabs.harvard.edu/abs/2022A&A...663A.123D}
}
\bibliographystyle{spiebib} 

\textcolor{blue}{
\paragraph{ Biographies \\ }
}

\textbf{Emmanuel Proven-Adzri} is an astrophysicist and Project Lead at the Ghana Radio Astronomy Observatory. With more than a decade of experience in astronomy and space science development, his work bridges instrumentation, computational modelling, and machine learning. Passionate about innovation and capacity building, he plays a leading role in advancing Africa’s presence in global radio astronomy while inspiring the next generation of scientists and technologists. \\

\textbf{Nia Imara} is an astrophysicist, artist, and programs director for the University of California–Santa Cruz GRAO Astronomy Development Project. The first Black woman to earn a PhD in astrophysics from UC Berkeley, she completed postdoctoral research at Harvard and now studies star formation in the Milky Way and other galaxies. She also founded Onaketa, providing free STEM tutoring and educational resources to Black and Brown youth.\\

\textbf{Theophilus Ansah-Narh} is a Senior Research Scientist and Acting Deputy Director at the Ghana Space Science and Technology Institute (GSSTI). His research covers radio astronomy, intensity mapping, and data calibration for radio telescopes, focusing on correcting instrumental effects. He has contributed to initiatives such as DARA and the Baby Telescope projects, advancing Ghana’s radio astronomy infrastructure and capacity-building while applying machine learning and data analytics to space science challenges across Africa.\\

\textbf{Wonder Sewavi} is a graduate student (MPhil) at the Department of Physics, KNUST. His work focuses on single-dish monitoring, with
specific interest in high-mass star formation and physical properties driving periodic variability in masers. He is also passionate about AGN, pulsars and numerical simulation, aiming to deepen our understanding of the cosmos through observation and modeling. His goal is to contribute to our understanding of star-forming regions and galactic evolution. \\

\textbf{Pieter Pretorius} is an Electronic Engineering graduate (B Ing, University of Stellenbosch, South Africa), currently Lead Software Engineer at SARAO (South African Radio Astronomy Observatory). He specializes in embedded and server-based systems, radar, communications, signal processing, and control systems. Currently leading software development converting a telecommunications antenna into Ghana’s Radio Astronomy Telescope. Outside work, he enjoys reading, coffee, and lifelong learning—always exploring new ideas and pushing boundaries through innovative technology.\\

\textbf{Evaristus U. Iyida} is an astrophysicist specializing in multi-wavelength studies of blazars and radio galaxies. He obtained his Ph.D. in 2022 from the University of Nigeria. He is currently a postdoctoral fellow at the Ghana Radio Astronomy Observatory, Ghana Space Science and Technology Institute, where he investigates radio continuum sources and their variability using Ghana’s 32-meter radio telescope.\\

\textbf{Naomi Asabre Frimpong} is an astrophysicist and science communicator based at the Ghana Space Science and Technology Institute (GSSTI). She holds a PhD in Astronomy/Astrophysics from the University of Manchester, UK, with research focusing on the astrochemistry of massive young stellar objects. Dr. Asabre Frimpong leads science communication and outreach at GSSTI and is currently the Deputy Director of the International Astronomical Union's Office for Astronomy Outreach. \\

\textbf{Joseph Bremang Tandoh} is the Director of the Ghana Space Science and Technology Institute (GSSTI). His leadership focuses on advancing Ghana’s capabilities in space science, satellite technology, and astronomy. He oversees research and development programs that strengthen national infrastructure, promote innovation, and build scientific capacity. Dr. Tandoh also fosters international collaborations, guiding strategic initiatives that position Ghana as a key contributor to Africa’s growing space science landscape.\\

\textbf{Diana Klutse} is an astrophysicist whose research focuses on developing and implementing advanced machine learning pipelines for the calibration of radio astronomy data at the Ghana Radio Astronomy Observatory (GRAO). She obtained her Ph.D. in 2025 from the University of KwaZulu-Natal (UKZN), South Africa, and currently serves as a postdoctoral fellow at the Ghana Space Science and Technology Institute, advancing data-driven approaches in astronomy.\\

\textbf{Benedicta Woode} is a research scientist at the Ghana Space Science and Technology Institute (GSSTI) under the Ghana Atomic Energy Commission and an adjunct lecturer with the Development in Africa with Radio Astronomy (DARA) program. Her research focuses on the formation of massive stars, using single-dish and Very Long Baseline Interferometry (VLBI) techniques to study masers and continuum emission. She leads the science commissioning of Ghana’s 32-metre radio telescope at Kuntunse and contributes to regional capacity building in radio astronomy. Her broader interests include telescope operations, data calibration, and space science education across Africa.\\

\begin{table}[H]
\centering
\caption{Technical Specifications of the GRAO 32\,m Telescope}
\label{tab:grao_spec}
\scriptsize  
\setlength{\tabcolsep}{6pt}  
\resizebox{\textwidth}{!}{
\begin{tabular}{ |l|l| }
\hline
\textbf{Category} & \textbf{Specification} \\
\hline
\multicolumn{2}{|c|}{\textbf{Location}} \\
\hline
Latitude/Longitude & \SI{5.750485}{\degree N}, \SI{-0.305116}{\degree W} \\
\hline
\multicolumn{2}{|c|}{\textbf{Antenna Structure}} \\
\hline
Mount Type & Alt-azimuth (wheel-and-track) \\
Optics & Shaped Cassegrain with beam-waveguide (2 concave, 2 flat mirrors) \\
Diameter & \SI{32}{\meter} \\
Height (zenith/horizon) & \SI{38.3}{\meter} / \SI{37.5}{\meter} \\
Movable Mass & \SI{230}{\tonne} \\
Azimuth Range & $\pm$\SI{300}{\degree} (relative to North) \\
Elevation Range & \SI{7}{\degree} to \SI{90}{\degree} \\
Surface Accuracy (RMS) & \SI{1.88}{\milli\meter} \\
Main Reflector Panels & 240 (thickness: \SI{1.6}{\milli\meter}) \\
\hline
\multicolumn{2}{|c|}{\textbf{Performance Parameters}} \\
\hline
Max Gain & \SI{63.72}{\dB} (5 GHz), \SI{66.40}{\dB} (6.7 GHz) \\
Aperture Efficiency & 81\% (5 GHz), 77\% (6.7 GHz)\\
Half-Power Beamwidth (HPBW) & \SI{0.11}{\degree} (5 GHz), \SI{0.09}{\degree} (6.7 GHz)\\
Sidelobe Level & \SI{-15.21}{\dB} (5 GHz), \SI{-15.15}{\dB} (6.7 GHz)\\
Spillover Loss & \SI{0.19}{\dB} (5 GHz), \SI{0.14}{\dB} (6.7 GHz)\\
Cross-Pol Isolation & \SI{-31.88}{\dB} (5 GHz), \SI{-32.21}{\dB} (6.7 GHz)\\
System Temperature ($T_{\text{sys}}$) & \SI{125}{\kelvin} (5 GHz), \SI{<90}{\kelvin} (6.7 GHz)\\
Sensitivity ($A_{\text{eff}}/T_{\text{sys}}$) & 
\SI{5.21}{\square\meter\per\kelvin} (5 GHz), 
\SI{4.95}{\square\meter\per\kelvin} (6.7 GHz)\\
\hline
\multicolumn{2}{|c|}{\textbf{Receiver System}} \\
\hline
Polarization & Dual circular (L + R) \\
Cooling & Ambient (uncooled) \\
Noise Temperature & \SI{110}{\kelvin} (5 GHz), \SI{<90}{\kelvin} (6.7 GHz) \\
Local Oscillator Tuning & \SI{1}{\hertz} resolution \\
Frequency Range & \SI{4.917}{}--\SI{5.045}{\GHz} (5 GHz band), \SI{6.55}{}--\SI{6.95}{\GHz} (6.7 GHz band) \\
\hline
\multicolumn{2}{|c|}{\textbf{Dynamic Performance}} \\
\hline
Slew Rate (Azimuth/Elevation) & \SI{0.27}{\degree\per\second} / \SI{0.29}{\degree\per\second} \\
Acceleration (Azimuth/Elevation) & \SI{0.056}{\degree\per\square\second} / \SI{0.072}{\degree\per\square\second} \\
Tracking Accuracy & \SI{0.05}{\degree} RMS \\
\hline
\end{tabular}%
}
\end{table}


\begin{table}
\centering
\caption{Expected S/N for the Vela pulsar at 5 GHz for a 61 minute integration and $SEFD$=406 Jy.}
\label{tab:vela_snr}
\begin{tabular}{lcc}
\hline
Bandwidth (MHz) &  Expected S/N \\
\hline
10   & 2.9 \\
50   & 6.5\\
100  & 9.2 \\
200  & 13.0\\
400  & 18.4 \\
800  & 26.0 \\
\hline
\end{tabular}
\end{table}

\textcolor{blue}{
\paragraph{ Figure Captions \\ }}

Figure1: The Ghana Radio Astronomy Observatory\\

Figure 2: Antenna performance simulation for the GRAO.
    Left panel: Normalized beam gain patterns at $5.0$ GHz and $6.7$ GHz, including sidelobe specifications (dashed lines). The simulations assume ideal optical alignment and a surface RMS error of $1.88$ mm.
    Right panel: Aperture efficiency as a function of elevation angle, incorporating surface deformation, atmospheric opacity, and pointing losses. Dashed vertical lines denote typical operational elevations ($30^\circ$, $45^\circ$, and $70^\circ$). Peak efficiencies match design specifications at zenith, while degradation at low elevation reflects expected structural and atmospheric constraints. \\

Figure 3: Methanol maser observation towards Cepheus A HW2 with the Kutunse radio telescope in Ghana on 04 March 2025. The axes are labeled for flux density (Jy) and Velocity of local standard of rest. \\

Figure 4: A 61-minute pulse profile of PSR J0835-4510 Vela pulsar taken with the GRAO. \\

Figure 5: VLBI fringe tests to the GRAO telescope (Kutunse) on baselines to the telescopes in Medicina, Yebes, Zelenchukskaya and Hartebeesthoek\cite{Gurvits2021}

\end{document}